**Resonant photoluminescence and excitation spectroscopy of CdSe/ZnSe and CdTe/ZnTe self-assembled quantum dots**

T. A. Nguyen*, S. Mackowski, H. Rho, H. E. Jackson, L. M. Smith
Department of Physics, University of Cincinnati, OH 45221-0011, United States
* corresponding author, electronic mail: tuanna@physics.uc.edu
J. Wrobel, K. Fronc, J. Kossut, and G. Karczewski
Institute of Physics, Polish Academy of Science, Warsaw, Poland.
M. Dobrowolska and J. Furdyna
Department of Physics, University of Notre Dame, United States

## ABSTRACT

We show that two major carrier excitation mechanisms are present in II-VI self-assembled quantum dots. The first one is related to direct excited state – ground state transition. It manifests itself by the presence of sharp and intense lines in the excitation spectrum measured from single quantum dots. Apart from these lines, we also observe up to four much broader excitation lines. The energy spacing between these lines indicates that they are associated with absorption related to longitudinal optical phonons. By analyzing resonantly excited photoluminescence spectra, we are able to separate the contributions from these two mechanisms. In the case of CdTe dots, the excited state – ground state relaxation is important for all dots in ensemble, while phonon-assisted processes are dominant for the dots with smaller lateral size.

## INTRODUCTION

The results of optical spectroscopy reported for III-V semiconductor self-assembled quantum dots (QDs) show that both excited states as well as electron-phonon interaction play an important role in carrier excitation and relaxation in these structures [1-4]. However, due to relatively weak confinement, it is sometimes hard to distinguish between these two processes, especially if the ground state – excited state energy interval equals to a multiple of the optical phonon energy [2].

In this work we investigate excitation and relaxation mechanisms in CdTe and CdSe QDs. As reported by several groups, QDs made of II-VI compounds are usually much smaller than InAs/GaAs structures [5-7]. This leads to stronger electronic confinement which, in turn, should give larger value of excited state – ground state splitting. Moreover, due to the different growth mechanism, no wetting layer formation is observed in the case of fully developed II-VI QDs [5.6]. The results of resonantly excited photoluminescence (PL) together with photoluminescence excitation (PLE) data obtained for single QDs feature unambiguously both types of mechanisms mentioned above. In particular, we show that the excited state (ES) – ground state (GS) processes are important for all QDs. In contrast, optical phonon assisted absorption is dominant only for QDs with high emission energy, i.e. presumably with smaller size. This finding agrees with recently predicted size dependence of the strength of electron – phonon coupling in QDs [8].

## SAMPLES AND EXPERIMENTAL DETAILS

The samples studied were grown by molecular beam epitaxy on GaAs substrate. After a 1μm thick ZnTe or ZnSe buffer layer 4 monolayers of CdTe or 2.6 monolayers of CdSe, respectively were deposited to form a random distribution of quantum dots. The dot layer was then covered by 50 nm thick ZnTe or ZnSe capping layers. Further details of the samples growth can be found elsewhere [5,6]. We point out here that unlike III-V QDs, no wetting layer is present in the structures studied here.

The QDs were excited either non-resonantly by the 514 nm line of an Argon-ion laser, or resonantly by a tunable dye laser (Rhodamine 590 and Coumarin for CdTe and CdSe, respectively). The sample was placed in a helium-vapor cooled cryostat with the temperature ranging from 6 K to 80 K. The laser was focused onto the sample surface with a spot diameter between 10 μm and 1 μm. The emitted PL was dispersed by a DILOR XY triple spectrometer working in a subtractive mode and was collected by $LN_2$-cooled CCD detector. The micro-PLE measurements were performed only for CdTe QDs. In this case the emission was excited through an aperture in an opaque metal mask which further reduces the number of QDs studied. The micro-PLE spectra were normalized to the laser power. The spectral resolution for both PL and micro-PLE measurements was about 100 μeV.

## RESULTS AND DISCUSSION

In Figure 1 we present a typical PLE spectrum obtained from a single CdTe/ZnTe QD collected through an aperture with a diameter of 0.8 μm. Two different types of resonances are observed in the spectra. The excitations with relatively weak intensity and with a linewidth of about 6 meV are associated with longitudinal optical (LO) phonon assisted absorption.

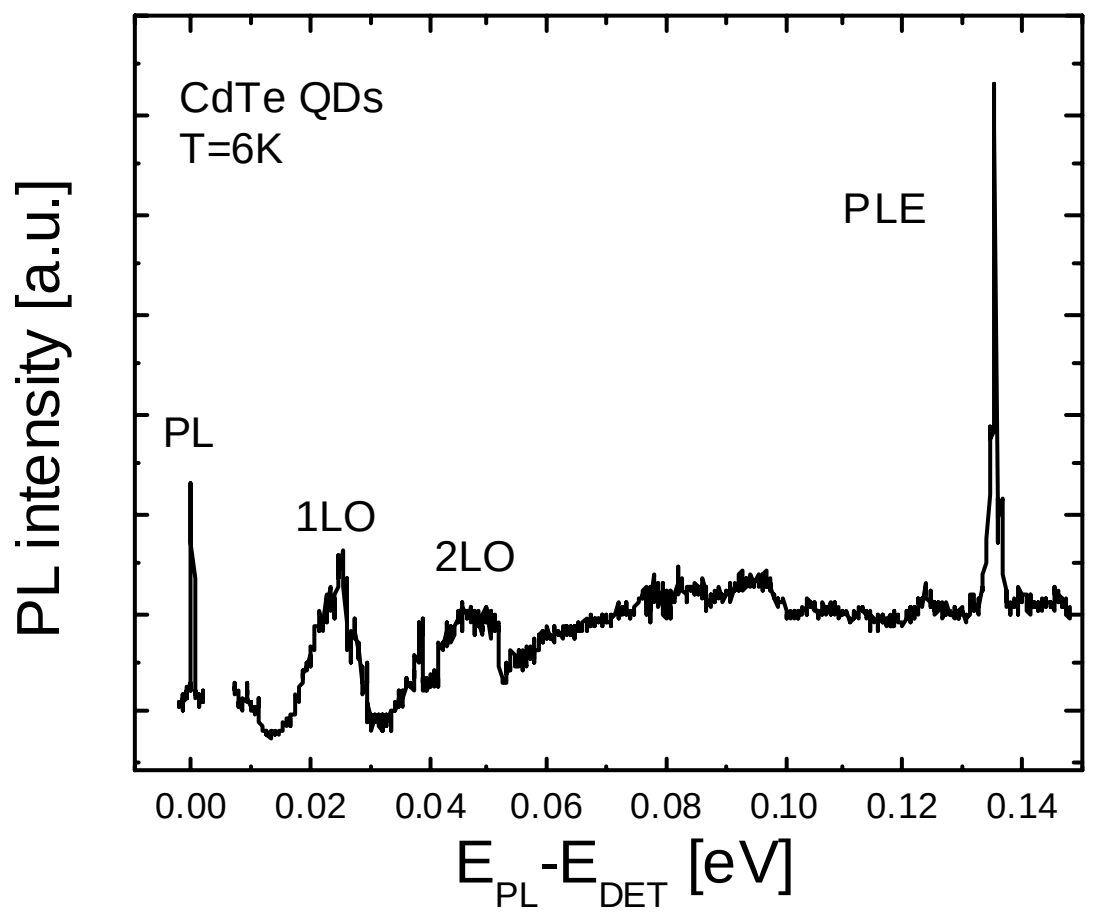

**Figure 1:** Typical low temperature (T=6 K) PLE spectrum obtained for a single CdTe/ZnTe QD collected through an aperture with a diameter of 0.8 μm. Both LO phonon – related and excited state – related resonances are shown.

Up to three such resonances are observed for CdTe QDs. All of them maintain the shape of the first LO absorption line while their intensity decreases for each successive replica. Another type of resonance visible in Figure 1 is characterized by much sharper lines which usually are also more intense. Both the energy difference and the intensity of these sharp lines depend on the particular QD. We tentatively assign them to the direct excited state (ES) – ground state (GS) excitations. The variation of the energy of these excitations reflects the size and/or composition fluctuations, which are commonly present in a QD ensemble. We find that on average the energy splitting between excited and ground state in CdTe QDs seems to be larger than for other QDs systems [9,10]. Since in a good approximation ES-GS energy distance increases with increasing carrier localization, the values of the splitting obtained for CdTe QDs would presumably suggests that a strong spatial confinement is present in this structure. This is additionally supported by a very small size of the structures (lateral diameter ~ 3nm), as evidenced by both structural and magneto-optical experiments [11,12].

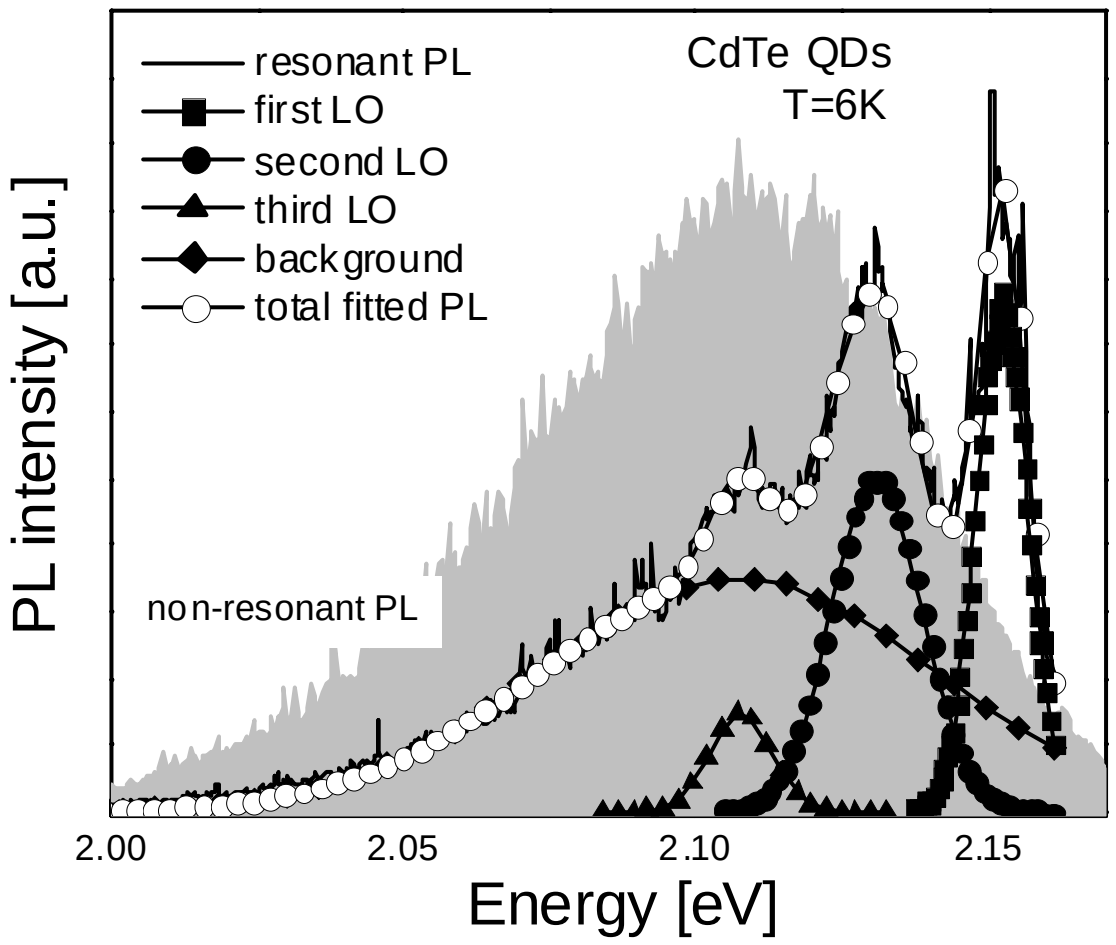

**Figure 2:** Resonantly excited PL spectrum (solid line) obtained for CdTe QDs compared with the non-resonant one (gray area). The open points show the fit to the spectrum. Full symbols are the first (squares), second (circles) and third (triangles) LO phonon replicas and the background luminescence (diamonds). The open symbols give the sum of all components.

On the basis of micro-PLE results one may expect that when probing a larger QD ensemble it would be possible to observe both LO phonon assisted processes as well as direct ES - GS relaxation. In Figure 2 we present the PL spectra obtained at T=6 K by exciting resonantly the QDs-related emission band. In this case, the PL emission is composed of up to four clearly separated lines superimposed on the relatively broad background. As for micro-PLE spectra, these lines represent exciton – LO phonon relaxation processes in QDs. However, in contrast to the micro-PLE, it seems that the spectral features related to LO phonon enhanced PL are becoming broader for the second and each successive replica, when compared to the first one. This broadening may be attributed to acoustic-phonon scattering, which could become as efficient as higher order optical phonon relaxation processes.

In order to analyze the excitation energy dependence of both the LO phonon resonances and the background PL intensities we fit each spectrum with series of Gaussian lines. All fits were performed under the assumption that the energy and linewidth of non-resonantly excited QDs PL spectrum characterize the background emission. In other words, it reflects the ground energy distribution within the QD ensemble. A typical result of the fitting is shown in Figure 2. The extracted integrated intensities of the first, second and third LO phonon replicas together with the background emission are plotted in Figure 3 as a function of the excitation energy. The non-resonantly excited PL spectrum is also shown for comparison. The results the LO phonon – related intensity variations with excitation energy are considerably narrower than the non-resonant spectrum. On the other hand, the width of the excitation energy dependence of the background PL (solid line in Figure 3) is comparable to the linewidth of non-resonant PL.

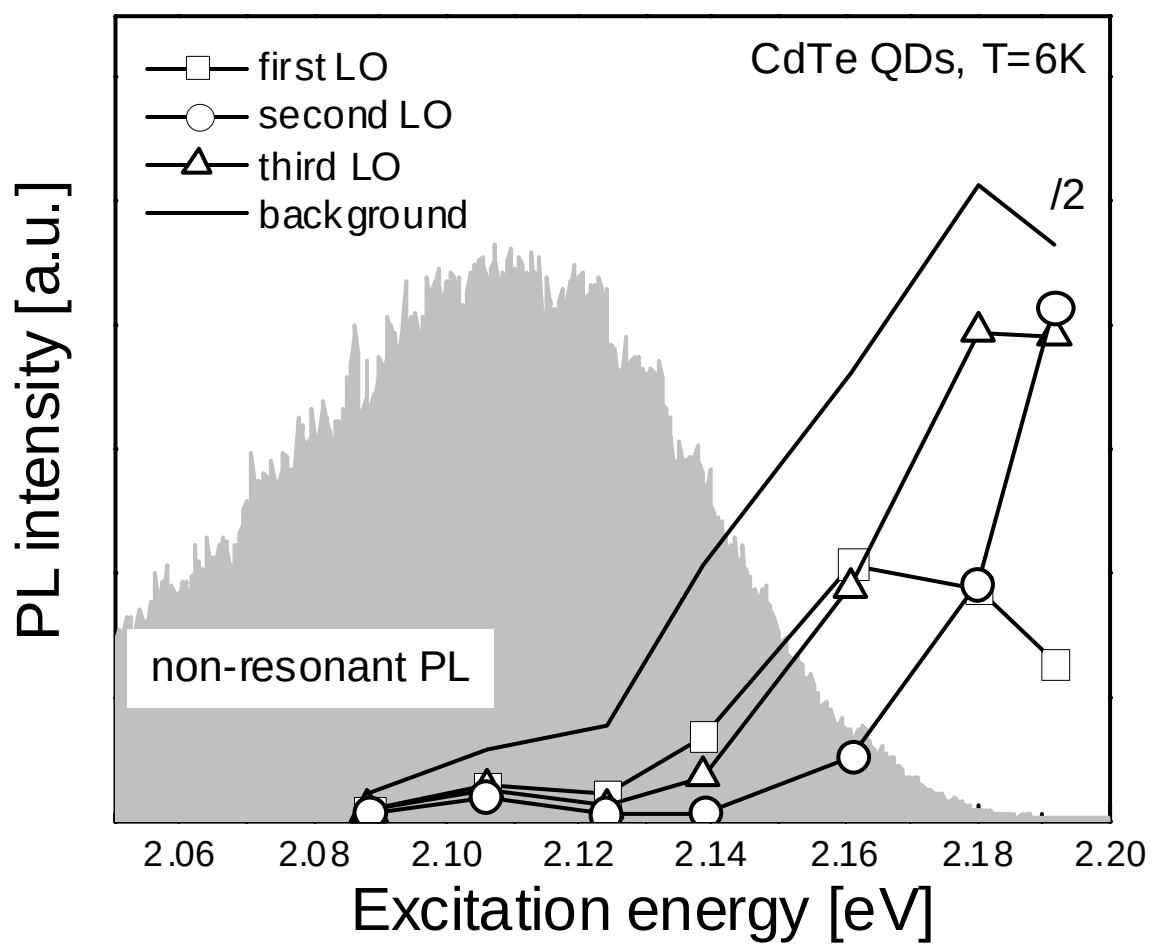

**Figure 3:** Integrated intensity of first LO (squares), second LO (circles), third (triangles) and the background (solid line) luminescence obtained for CdTe QDs as a function of excitation energy. The non-resonant spectrum is also shown for comparison.

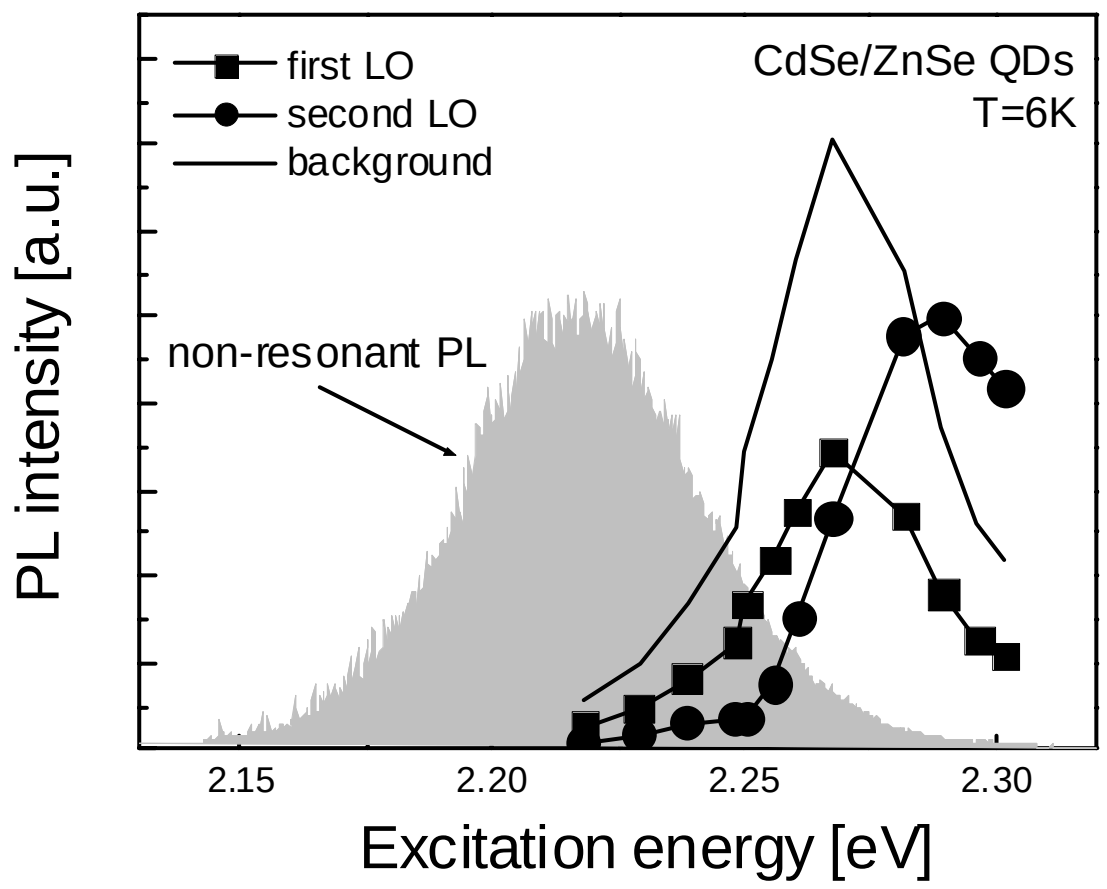

**Figure 4:** Integrated intensity of first LO (squares), second LO (circles) and the background luminescence obtained for CdSe QDs as a function of excitation energy. The non-resonant spectrum measured for this sample is also shown for comparison.

Similar measurements have been also performed on QDs made of another II-VI semiconductor system, namely CdSe/ZnSe. The results of the same analysis as described previously are presented in Figure 4. In this case the excitation energy dependence of all spectral features (first and second LO phonon replicas and the background) seem to mimic the shape of non-resonant PL spectrum. Additionally, the maximum of the first LO phonon replica occurs approximately 30 meV above the non-resonant PL which agrees approximately with an average LO phonon energy in this system. A different situation is observed for CdTe QDs, where the first LO phonon resonance is found more than 60 meV above the maximum of non-resonantly excited PL.

The resonant PL spectroscopy results can be explained in the following way. We assume that the non-resonant spectrum reflects the true GS energy distribution of the QD ensemble. In both QDs systems studied, the background luminescence reflects then the GS energy distribution in a particular system. The energy spacing between non-resonant PL and the maximum enhancement of background PL is very similar to the energy range where the sharp intense lines are observed in micro-PLE spectra for CdTe QDs. The same kind of correspondence is found for CdSe QDs. Indeed the energy spacing of around 50 meV estimated from Figure 3 agrees well with micro-PLE results published recently for CdSe QDs of the lateral size very similar to the structures studied here [9]. Indeed, in this work [9] a series of sharp lines within the range of 40 meV – 50 meV above the single dot GS energy was observed. All of these observations indicate that both sharp lines in micro-PLE spectra and the background emission in resonantly excited PL experiments are related to direct relaxation from an excited QD level down to its ground state.

While the excitation energy dependence of the background emission is qualitatively identical for both CdTe and CdSe QDs, the LO phonon-assisted processes show some important differences. A possible explanation for observed differences could be associated with the QDs size dependence of the exciton – phonon coupling. Indeed, as suggested by recent theoretical calculations, the strength of exciton – phonon coupling features a pronounced maximum for QDs with a diameter of about 2 nm [8], while for larger dots this dependence is relatively weak. The latter is the case for CdSe QDs, where both first and second LO phonon behavior as a function of excitation energy reflects the distribution of GS QD energy. Moreover, the energy separation between the maximum enhancement of LO phonon replica and the non-resonant PL is equal to an average LO phonon energy in this system. This suggests that the LO phonon-assisted processes in CdSe QDs are governed predominantly by the density of states distribution, in the same sense as the direct ES – GS relaxation processes discussed earlier.

On the other hand, the experimentally observed behavior for CdTe QDs indicates that excitons in certain dots couple to LO phonons much strongly (cf. Figure 2). First of all, the linewidth of the excitation energy dependence obtained for the first LO phonon replica is much narrower than the QD energy distribution. Secondly, the energy separation between non-resonant PL maximum and that of the first LO phonon enhancement does not correspond to any multiple of an average phonon energy in the system. Thus, we suggest that the maximum we observe in the excitation energy dependence of LO phonon enhancement reflects the QD size dependence of the exciton – phonon coupling in zero-dimensional structure. Moreover, the average size of CdTe QDs studied here is 3 nm [12], which agrees reasonably with the one predicted to show the largest exciton - phonon coupling [8].

## CONCLUSIONS

We show that two major carrier excitation mechanisms are present in CdTe and CdSe QDs. The first one is a direct excited state – ground state relaxation and is predominantly governed by the ground state energy distribution within the QD ensemble. The second one is associated with an LO phonon-assisted absorption. By analyzing PL spectra obtained under resonant excitation, we show that the strength of exciton – phonon coupling features a maximum for dots with small lateral size (CdTe). On the other hand, for larger QDs (CdSe) the intensities of LO phonon replicas reflect the ground state energy distribution in the QD ensemble.

## ACKNOWLEDGEMENTS

We gratefully acknowledge the financial support of the National Science Foundation through grants NSF DMR 0071797, 9975655 and 0072897 and the DARPA-SPINS Program. Work in Poland supported by PBZ-KBN-044/P03/2001.